\documentclass[prl,twocolumn,groupeaddress]{revtex4-2}

\usepackage{amsmath}
\usepackage{graphicx}
\usepackage[colorlinks=true, allcolors=blue]{hyperref}
\usepackage{bbold}

\begin{document}

\title{Interplay of orbital-selective Mott criticality and flat-band physics in La$_3$Ni$_2$O$_6$}
\author{Frank Lechermann}
\affiliation{Theoretische Physik III, Ruhr-Universit\"at Bochum,
  D-44780 Bochum, Germany}
\author{Steffen B\"otzel}
  \affiliation{Theoretische Physik III, Ruhr-Universit\"at Bochum,
  D-44780 Bochum, Germany}
\author{Ilya M. Eremin}
\affiliation{Theoretische Physik III, Ruhr-Universit\"at Bochum,
  D-44780 Bochum, Germany}

\begin{abstract}
  Superconductivity in nickelates apparently takes place in two different Ni oxidation regimes,
  namely either for infinite-layer-type compounds close to Ni$^{+}$, or for Ruddlesden-Popper materials
  close to Ni$^{2+}$. The reduced La$_3$Ni$_2$O$_6$ bilayer with a nominal Ni$^{1.5+}$ oxidation state may
  therefore serve as a normal-state mediator between the two known families of $3d^8$-like and $3d^9$-like
  superconducting nickelates. Using first-principles many-body theory, we explain its experimental 55\,meV
  charge gap as originating from a new type of correlated (quasi-)insulator. Flat-band electrons of
  Ni-$d_{z^2}$ character become localized from scattering with orbital-selective Mott-localized Ni-$d_{x^2-y^2}$
  electrons, by trading in residual hopping energy for a gain in local exchange energy in a
  ferromagnetic Kondo-lattice scenario. Most importantly, the flat-band electrons offer another route to
  unconventional superconductivity in nickelates at ambient pressure. 
\end{abstract}

\maketitle
\textit{Introduction.---}
The long-standing research on nickel oxides has gained enormous new momentum in recent years
because of the findings of superconductivity in two different nickelate families.
First, in Ni$(3d^{9-\delta})$ reduced materials, as originally given by thin films of Sr-doped
infinite-layer NdNiO$_2$ with a $T_{\rm c}\sim 15$\,K~\cite{li19}. And second, in Ni$(3d^{8\pm\delta})$
Ruddlesden-Popper (RP) materials, as started off by the discovery of a $T_{\rm c}\sim 80$\,K in
bulk bilayer La$_3$Ni$_2$O$_7$ at pressures $p>14$\,GPa~\cite{sun23}.
Correlated electronic processes in the $e_g$ subshell of Ni$(3d)$, built up by
the $\{d_{x^2-y^2},d_{z^2}\}$ orbitals, are most likely at the root of superconductivity. Yet
the respective role of both $e_g$ orbitals still needs deeper understanding in both
families, i.e. reduced nickelates~\cite{wu19,zhangzhang20,werner20,lechermann20-1,karp20,leonov20,adhikary20,sakakibara20,Kitatani2020,been21,geisler21,kang21,gu2020substantial,plienbumrung22,jiang22,foyevtsova23}
and RP nickelates~\cite{luohu23,lechermann23,zhanglin23,lechermann23,shilenko23,chen-jul23,sakakibara24,christiansson23,yangwangwang23,lupan23,liumei23,yangzhang23,xing-zhou24,jianghuo23,luobiao23,oh24,boetzel24,ryee24,Savrasov2024}.
But the existence of two seemingly different superconducting families is
fascinating and goes beyond the single-family paradigm of Cu$(3d^{9\pm\delta})$ high-$T_{\rm c}$ cuprates.

\begin{figure}[t]
      \includegraphics[width=\linewidth]{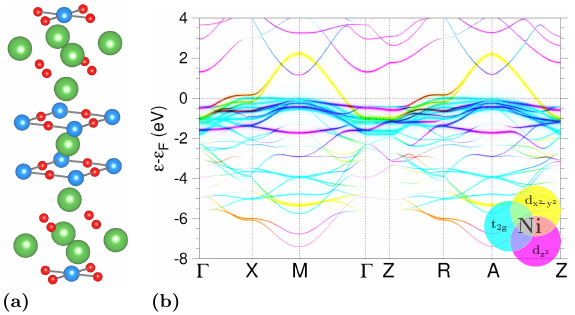}
      \caption{Reduced bilayer compound La$_3$Ni$_2$O$_6$. (a) Tetragonal 
      $I4/mmm$ crystal structure with La(green), Ni(blue) and O(red) atoms. (b) DFT dispersions in fatspec representation highlighting the Ni$(3d)$ orbital characters.
}\label{fig1}
\end{figure}

In this context, the La$_3$Ni$_2$O$_6$ compound (see Fig.~\ref{fig1}a) stands out and may attain a
unique role in connecting both superconducting nickelate families.
As the fully reduced form of La$_3$Ni$_2$O$_7$, it shares the bilayer structure but also
displays the missing apical oxygens of the original reduced systems. Furthermore, its nominal
Ni$^{1.5+}$ oxidation state with $3d^{8.5}$ electron count just mediates between the latter $3d^9$-like
systems and the ligand-hole affected $3d^8$-like RP compounds. Last but not least, it crystallizes
in tetragonal symmetry, which is the common symmetry for superconducting nickelates.

Experimentally, the material has first been reported in 2006 in powder form~\cite{poltavets06}, and
soon afterwards characterized as semiconducting with no magnetic order down to 4\,K~\cite{poltavets09}.
Absence of both, static order and metallicity, has also been concluded from nuclear magnetic
resonance experiments~\cite{roberts-warren13}. One may attribute semiconductivity to
the polycrystalline samples, yet recent single-crystal studies confirmed non-metallicity,
revealing a small activation gap of 55\,meV at low $T$~\cite{liu23}. Magnetization measurements
show a weak kink at $T~\sim 176$\,K, but otherwise again no obvious signature
of a magnetic transition.

Early density functional theory (DFT) plus Hubbard U calculations resulted in a metallic
state for La$_3$Ni$_2$O$_6$~\cite{poltavets09}. Additional follow-up DFT+U investigations predicted
an intriguing charge-ordered Ni$^{+}(S=1/2)$/Ni$^{2+}(S=0)$ insulating ground state with magnetic
order~\cite{botana16}. Albeit the theoretical charge gap of 550\,meV is an order of magnitude
larger than the eventually measured gap. For high pressure, a very recent DFT+U study finds a low-spin ferromagnetic order as the likely ground state~\cite{zhanglin24}.
\begin{figure*}[t]
      \includegraphics[width=\linewidth]{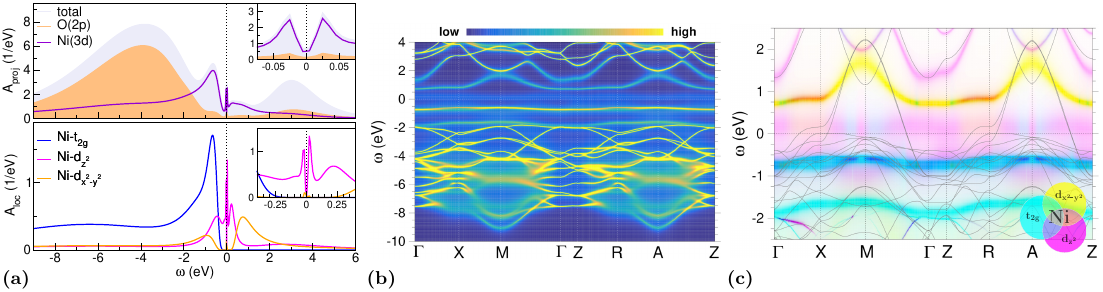}
      \caption{Spectral properties of La$_3$Ni$_2$O$_6$ from DFT+sicDMFT at $T=50$\,K.
        (a) Upper panel: total and projected {\bf k}-integrated
        spectral function (inset: low-energy blow up). Lower panel: Ni$(3d)$ local spectral function. Right inset: low-energy blow up.
        (b) {\bf k}-resolved spectral function $A({\bf k},\omega)$ along
        high-symmetry lines. (c) Fatspec representation of $A({\bf k},\omega)$ for the
        different Ni$(3d)$ contributions (DFT bands: grey lines).} \label{fig2}
\end{figure*}	

In this work, we report a theoretical description of the correlated electronic structure of
paramagnetic La$_3$Ni$_2$O$_6$ based on first-principles many-body theory. The experimental 55\,meV
gap is well reproduced, being the result of a novel correlated insulating scenario. It builds up
on orbital-selective Mott physics~\cite{goodenough82,anisimov02} of localized Ni-$d_{x^2-y^2}$ electrons
acting on a Ni-$d_{z^2}$ flat-band feature. Interestingly, the flat-band low-energy dispersion also
becomes gapped and the emerging semiconducting state is rather robust against hole doping $x$. It is
only metallized for $x>0.15$ when the flat-band electrons are released from (quasi-)localization.
The strong flat-band response at low-energy with threshold hole doping may signal further electronic
instabilities including unconventional superconductivity with sign-changing $s_{\pm}$-wave symmetry.

\textit{Correlated electrons.---} Previously, a realistic dynamical mean-field theory (DMFT) study of
La$_3$Ni$_2$O$_7$~\cite{lechermann23} revealed half-filled Ni-$d_{x^2-y^2}$ and Ni-$d_{z^2}$ orbitals,
respectively. Low-energy electrons in both orbitals are subject to strong correlations,
yet the Ni-$d_{z^2}$ dispersion has substantial flat-band character at the Fermi level.
As a result, the highly correlated metallic state is prone to instabilities, including
superconductivity at high pressure.
For electrons in the reduced bilayer La$_3$Ni$_2$O$_6$ we also expect strong correlations,
however, with much larger orbital differentiation due to the missing apical oxygens. 
As shown for infinite-layer NdNiO$_2$ using the same theoretical framework~\cite{lechermann20-1},
Ni-$d_{x^2-y^2}$ is even close to Mott localization, whereas ``free-standing'' Ni-$d_{z^2}$ turns
out weakly correlated. Yet from formal $d^9$ in NdNiO$_2$ the Ni-$d_{z^2}$ electron count is near
complete filling, whereas from formal $d^{8.5}$ in La$_3$Ni$_2$O$_6$ a much lower filling is expected. 

The La$_3$Ni$_2$O$_6$ inspection on the DFT level (see Fig.~\ref{fig1}b) exhibits flat
Ni-$d_{z^2}$ dispersions just below the Fermi level. But there is additional prominent low-energy
flat-band character close to the X point from dominant Ni-$d_{xz,yz}$ character~\cite{labollita22},
as also observed for Nd$_3$Ni$_2$O$_6$~\cite{worm22}.
To assess the role of correlations in the reduced bilayer, we discuss
its electronic structure as described by the combination of DFT, self-interaction correction (SIC)
and DMFT, i.e. within the DFT+sicDMFT approach~\cite{lechermann19} (see Supplemental Material with
Refs.~\onlinecite{grieger12,korner10,elsaesser90,lechermann02,mbpp_code,werner06,parcollet15,seth16,amadon08,anisimov93,lechermann22-2,choikutepov16,kanggwedmft25} for further details). In the upper panel of
Fig.~\ref{fig2}a the total {\bf k} integrated spectral function and the projected Ni$(3d)$ and
O$(2p)$ content is displayed. A sizable charge-transfer energy
$\Delta_{\rm ct}=\varepsilon_d-\varepsilon_p$, for respective band centers
$\varepsilon_{d,p}$ of Ni$(3d)$ and O$(2p)$, leads to a spectral separation of both dominant
chemical contributions. A broad O$(2p)$ weight is centered around $-4$\,eV,
and the Ni$(3d)$ weight dominantly peaks at $\sim -0.8$\,eV. The latter peak is primarily formed
by Ni-$t_{2g}$ character, as revealed by the local spectral function in the lower panel of
Fig.~\ref{fig2}a. Spectral integration leads to occupations $n_p=5.6$ and $n_d=8.3$, i.e. both lower
than expected from nominal O$^{2-}$ and Ni$^{1.5+}$. Hence there is a sizable ligand-hole character
on oxygen, and a Ni$(3d)$ count closer to the favorable $d^8$ occupation. This means that
either the Ni$(4s)$ or the La$(5d,6s)$ sub-shells are not fully oxidized. Note that while the total
filling is physically well-defined, orbital fillings always depend on the choice of local
functions. The substantial ligand-hole content for a nickelate with nominal $n_d>8$ is unexpected,
though note that La$_3$Ni$_2$O$_6$ is formally also located in an unusual $\Delta_{\rm ct}$
regime~\cite{lechermann-ox24}. The obtained fillings still do not come as a total surprise,
as an electride-like setting has e.g been put forward for infinite-layer
nickelates~\cite{foyevtsova23}.
\begin{figure}[b]
      \includegraphics[width=\linewidth]{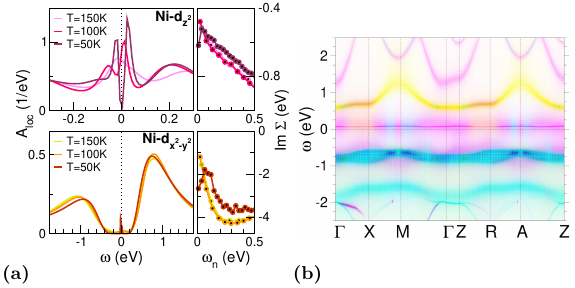}
      \caption{Effect of temperature on the Ni-$e_g$ states of La$_3$Ni$_2$O$_6$ from DFT+sicDMFT.
        (a) Temperature dependence of low-energy spectral function (left), and respective  
        imaginary part of the Matsubara self-energy $\Sigma(i\omega_n)$ (right). (b) Fatspec
        representation of the {\bf k}-resolved spectral function $A({\bf k},\omega)$ for $T=150$\,K.
}\label{fig3}
\end{figure}	
\begin{figure}[t]
      \includegraphics[width=\linewidth]{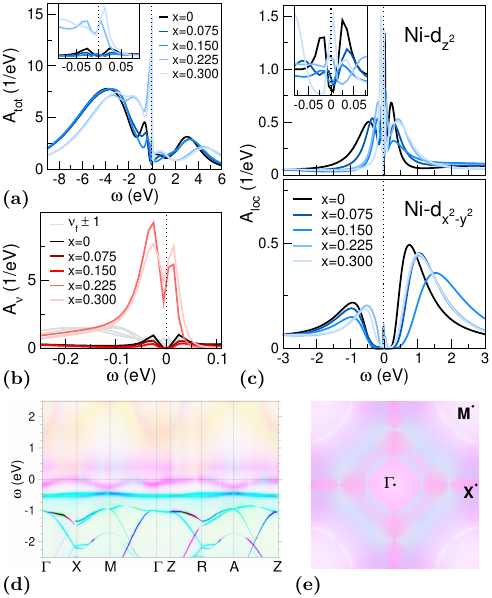}
      \caption{Spectral properties of hypothetical (La$_{1-x}$Sr$_x$)$_3$Ni$_2$O$_6$
        at $T=50$\,K based on DFT+sicDMFT using VCA for the doped La sites. (a) Total spectral function
        (inset: low-energy blow up). (b) Bloch-resolved spectral function $A_\nu(\omega)$
        for $\nu=\nu_{\rm f}$ (variants of red) and for $\nu=\nu_{\rm f}\pm 1$ (grey), with $\nu_{\rm f}$
        denoting the Bloch state, largest susceptible to hole doping.
        (c) Ni-$e_g$-resolved local spectral function
        (inset: low-energy blow up for Ni-$d_{z^2}$). (d,e) Fatspec representation of the
        {\bf k}-resolved spectral function $A({\bf k},\omega)$, and the Fermi surface in
        the $k_z=0$ plane for $x=0.225$. Color coding of (d,e) as in Fig.~\ref{fig2}c.
}\label{fig4}
\end{figure}	

At first glance, the low-energy spectral weight at $T=50$\,K seemingly shows a sharp
quasiparticle(QP)-like peak, but the detailed resolution (see inset of upper panel of Fig.~\ref{fig2}a)
displays a narrow (pseudo)gap feature of $\sim$50\,meV size, in very good agreement with the experimental
semiconducting gap of La$_3$Ni$_2$O$_6$~\cite{liu23}. From the local spectral function, 
this (pseudo)gap is exclusively linked to Ni-$d_{z^2}$, while Ni-$d_{x^2-y^2}$ is Mott-gapped
orbital selectively on a larger scale of $\sim 0.85$\,eV. Concerning the local-orbital fillings,
the value $n_{x^2-y^2}=1.10$ marks the expected near-half-filled scenario for Mott-gapped in-plane
$d_{x^2-y^2}$, whereas the higher value $n_{z^2}=1.27$ is realized for out-of-plane $d_{z^2}$.
As shown in the left panel of Fig.~\ref{fig3}a, the intriguing low-energy spectrum results from a
metal-to-insulator transition with lowering temperature. While the Ni-$d_{x^2-y^2}$ orbital is Mott-gapped
already well above 100\,K, the Ni-$d_{z^2}$ still displays metallic character with a seemingly QP-like
peak at $T=150$\,K. But in the spectrum at $T=100$\,K, the opening of a (pseudo)gap for the latter
orbital starts to become visible. For $T=150$\,K, Fig.~\ref{fig3}b indeed exhibits the semi-coherent
Ni-$d_{z^2}$ dominated flat band very close to the Fermi level.
Concomitantly, the imaginary part of the Matsubara self-energy $\Sigma(i\omega_n)$, shown in the right panel
of Fig.~\ref{fig3}a, looks Fermi-liquid like for Ni-$d_{z^2}$ over a wide frequency range.
Yet it bends downwards for smallest $\omega_n$ at $T=50$\,K, i.e. signaling the established (pseudo)gap
feature. The larger Ni-$d_{x^2-y^2}$ self-energy also changes qualitative behavior in the (pseudo)-gap regime,
pointing towards an intricate coupling scenario between both Ni-$e_g$ sectors.

Figures~\ref{fig2}b,c display the {\bf k}-resolved spectral function for La$_3$Ni$_2$O$_6$ at $T=50\,K$.
The original Ni-$t_{2g}$ DFT flat-band is repelled from the Fermi level by
correlations. Instead, the Ni-$d_{z^2}$ flat-band is shifted to lowest energy. This
correlated Ni-$d_{z^2}$ flat band becomes rather incoherent and splits, giving rise to the small
(pseudo)gap at low energy. Hence, flat-band physics as observed in La$_3$Ni$_2$O$_7$ is also present
in La$_3$Ni$_2$O$_6$, yet with a different twist. The substantial Ni-$e_g$ orbital anisotropy in
the reduced bilayer renders $d_{x^2-y^2}$ closer to Mottness, whereas $d_{z^2}$ is distant from Mott
criticality. But due to its low-energy flat-band electrons scattering with Mott-localized
$d_{x^2-y^2}$ electrons, it still becomes (pseudo)gapped. The question about an
intuitive explanation for this Ni-$d_{z^2}$ (pseudo)gapping arises. It should be based on the
competition between residual (flat-band) hopping and local-moment formation. Because of the already
localized Ni-$d_{x^2-y^2}$ electron with $S=1/2$, the system may gain exchange energy from hybridizing
with flat-band Ni-$d_{z^2}$ electrons. It has in fact been shown~\cite{biermann05}, that such a
two-band orbital-selective regime is described by a ferromagnetic Kondo-lattice Hamiltonian.
Non-Fermi-liquid behavior and pseudogap features are possible solutions of such
ferromagnetic/underscreened Kondo-lattice
problems~\cite{biermann05,giamarchi93,colemanpepin03,florens04}. Here, the
scattering between localized Ni-$d_{x^2-y^2}$ and flat-band Ni-$d_{z^2}$ leads to a
strong (pseudo)gap.
\begin{figure*}[t]
      \includegraphics[width=\linewidth]{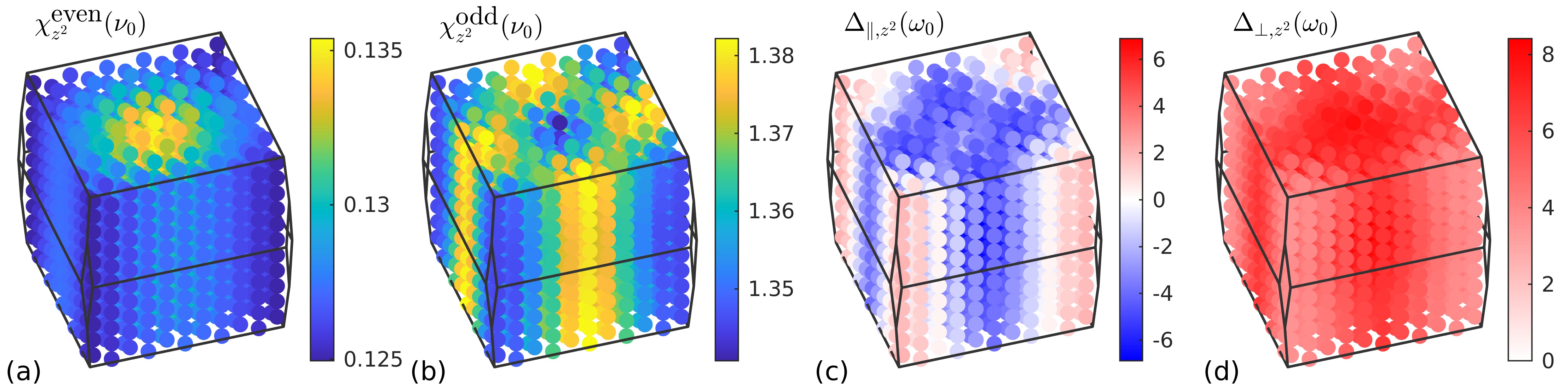}
      \caption{Spin susceptibility and superconducting gap function for the lowest positive Matsubara frequency of the Ni-$d_{z^2}$-orbital component in the first Brillouin zone. Note that the even susceptibility in (a) is an order of magnitude smaller than the odd susceptibility in (b). The real part of the intra- and interlayer superconducting gap functions are shown in (c) and (d), respectively. The susceptibility is given in units of states/eV, and the superconducting gaps are in units of meV.
}\label{fig5}
\end{figure*}	

\textit{Hole doping.---}
The behavior with doping now becomes of immediate interest. For instance, hole doping should lower
the Ni$(3d)$ electron count, eventually bringing it in line with the effective $d^8$ filling
of La$_3$Ni$_2$O$_7$. This could yield a novel correlated-electron setting for unconventional
superconductivity. Note that the application of pressure to stoichiometric
La$_3$Ni$_2$O$_6$ has already been performed up to 25\,GPa, resulting in metallicity but not
superconductivity~\cite{liu23}. This is reasonable, since pressure may close the (pseudo)gap because
of increasing hybridization. But this alone will likely not be sufficient to enable superconductivity,
since additionally also the filling has to be adjusted properly. While in high-pressure
La$_3$Ni$_2$O$_7$, the flat band is nearly fully occupied~\cite{lechermann23}, for present
stoichiometric La$_3$Ni$_2$O$_6$ the flat-band part seems not too far off half filling.
Even without pressure, hole doping alone will abolish the (pseudo)gapped state by effectively
raising the kinetic-energy gain above the exchange-energy gain for some doping level.

By employing the virtual-crystal approximation (VCA) for the doped La sites by Sr, we performed
DFT+sicDMFT calculations at hole doping levels $x$, and the results are depicted in
Fig.~\ref{fig4}.
From the total spectral function in Fig.~\ref{fig4}a, the (quasi-)insulating state
is robust up to a significant replacement of about 15\% of La by Sr. However, for
$x\gtrsim 0.2$ a strong QP resonance appears at the Fermi level. It is of dominant Ni-$d_{z^2}$
character as observed from the Ni-$e_g$-resolved local spectral functions in Fig.~\ref{fig4}c.
On the other hand, the Ni-$d_{x^2-y^2}$ electrons remain essentially localized throughout the 
studied doping range. While the orbital-selective Mott state is robust with hole doping,
the flat band is ``released'' from its (pseudo)gapped state with increasing $x$. The emerging
QP resonance is truly associated with the original flat-band state as displayed in Fig.~\ref{fig4}b.
There we plot the Bloch-resolved spectral function $A_\nu(\omega)$, i.e. the ${\bf k}$-integrated
interacting spectral function in the Bloch basis $\{\nu\}$, for the most susceptible $\nu=\nu_{\rm f}$.
One easily recognizes the large doping-dependent low-energy response. On the other hand, already
for interacting states of Bloch label $\nu_{\rm f}\pm 1$, there is only weak low-energy spectral
change with doping visible (grey lines in Fig.~\ref{fig4}b). This renders obvious that the
key correlation effects are local in Bloch space $\{\nu\}$, and with hole doping one unlocks
the dominant low-energy flat band dispersion from its (quasi-)insulating nature. This (still weak)
Ni-$d_{z^2}$ dispersion is finally seen in the {\bf k}-resolved spectral function of Fig.~\ref{fig4}d for
$x=0.225$, leading to electron pockets at X and hole pockets at M in the interacting Fermi surface,
as shown in Fig.~\ref{fig4}e.

\textit{Superconducting instability.---}
It is tempting to consider potential Cooper-pairing due to spin fluctuations in the
intriguing Ni-$d_{z^2}$ flat-band regime for hole doping $x=0.225$. A phase transition
to the superconducting state requires numerical evaluation of the pairing vertex, which consists of
the spin and charge susceptibilities and related irreducible vertices. We follow the
prescription developed previously for multiorbital systems~\cite{nourafkan2016correlation}, and
applied to La$_3$Ni$_2$O$_7$~\cite{ryee24,gao2024theoretical}, where the effective vertices are
independent of frequency and momentum. But here, the Green's functions are taken from the
converged DFT+sicDMFT calculation. For further details, we refer to the Supplemental Material. 

The spin and charge fluctuations are calculated using the Random Phase Approximation (RPA) for
both Ni-$e_g$ orbitals. However, the Ni-$d_{x^2-y^2}$ contributions are calculated to be
negligible, leaving the stage for Ni-$d_{z^2}$. In bilayer systems with only on-site interactions,
complete information requires calculating even and odd spin/charge susceptibilities
$\chi^{\rm even/odd} = \chi_\parallel \pm \chi_{\perp}$ \cite{yamase2024theory,boetzel24}. The former are
shown in Figs.~\ref{fig5}a,b, respectively, for an effective interaction corresponding to a Stoner
enhancement factor of $\alpha_{\rm sp} = 0.9$. A magnetic instability is indicated by
$\alpha_{\rm sp} = 1$. Note that the spin fluctuations in the odd channel are an order of magnitude
larger than in the even channel. The corresponding magnetic instability would thus have
antiferromagnetic interlayer intrabilayer order.  

The leading superconducting solution is a $s_\pm$-wave with large interlayer component, similar to
what is frequently discussed for La$_3$Ni$_2$O$_7$. The real part of the gap structure of the
leading instability is visualized in Figs.~\ref{fig5}c,d, focusing on the lowest Matsubara
frequency Ni-$d_{z^2}$ components. The imaginary part has a similar shape and is smaller in magnitude.
The shown $s_\pm$ solution is strongly dominant for all reasonable effective interactions and already
about unity for $\alpha_{\rm sp} = 0.9$, indicating a transition to the superconducting
state. Note that the interlayer component is not affected by on-site Coulomb repulsion, such that no
sign changes are required to avoid the repulsion. Furthermore, the combinations
$\Delta_\parallel \pm \Delta_\perp$ have opposite signs, which is also expected for the $s_\pm$ scenario
in La$_3$Ni$_2$O$_7$~\cite{boetzel24}. For the intralayer part, it is important to mention that there
is a non-trivial frequency dependence, with the higher-frequency components being dominated by the
opposite sign likely protecting the solution against Coulomb repulsion
(see Fig.~S1 in the Supplemental Material).  

For comparison, we also solved the frequency-dependent gap equation for pressurized La$_3$Ni$_2$O$_7$
using the DFT+sicDMFT Green's functions from Ref.~\cite{lechermann23} (see Supplemental Material for
details). There, the magnetic instability
for the correlated state is dominated by Ni-$d_{x^2-y^2}$. Different competing gap
structures of multiorbital and multilayer type, including not only a $s_\pm$-wave but also $d_{x^2-y^2}$
gap structure, have comparable eigenvalues. The eigenvalues approach unity only in close proximity to
the magnetic instability. Therefore, from the correlated DFT+sicDMFT picture, hole-doped
La$_3$Ni$_2$O$_6$ may even offer a better platform to realize strong-coupling unconventional
superconductivity of $s_\pm$ type with a dominant interlayer Ni-$d_{z^2}$ component than La$_3$Ni$_2$O$_7$.
At the same time, one has to bear in mind that other density-wave like instabilities, not included in
the present consideration, might affect this conclusion. 

\textit{Summary.---}
Using advanced first-principles many-body theory we identified the La$_3$Ni$_2$O$_6$
compound as a novel striking correlated material in the family of transition-metal oxides. The
reduced RP bilayer compound serves as a unique example where orbital-selective Mott and flat-band
physics join in to form a correlated (quasi-)insulator. It hosts orbital-selective Mott-localized
Ni-$d_{x^2-y^2}$ electrons and a flat-band electron state of
Ni-$d_{z^2}$ character which is gapped by $\sim$50\,meV. This gapping originates from scattering
between both orbital sectors within an anomalous Kondo-insulator scenario.
The reduced bilayer system thus serves as the long-sought explicit materials connection
between heavy-fermion and correlated-oxide physics.
Doping the flat-band state offers an interesting route for strong-coupling
unconventional superconductivity with a dominant $s_{\pm}$-wave symmetry solution
residing on Ni-$d_{z^2}$-bilayer splitted orbitals. This finding links the already established
superconducting regimes of RP- and infinite-layer-based nickelates. Last but not least, after the
recent discovery of superconducting La$_3$Ni$_2$O$_7$ under compressive
strain~\cite{Ko24,zhou-ambient24}, the present work paves the way for ambient-pressure
high-$T_{\rm c}$ superconductivity in bulk nickelates.

\section{Acknowledgements} 
The work is supported by the German Research Foundation within the bilateral NSFC-DFG Project ER 463/14-1. 
Computations were performed at the Ruhr-University Bochum and the JUWELS Cluster of the
J\"ulich Supercomputing Centre (JSC) under project miqs.

\bibliography{literatur}
\end{document}



\title{Supplemental Material: Interplay of orbital-selective Mott criticality and flat-band physics in La$_3$Ni$_2$O$_6$}
\author{Frank Lechermann}
\affiliation{Theoretische Physik III, Ruhr-Universit\"at Bochum,
  D-44780 Bochum, Germany}
\author{Steffen B\"otzel}
  \affiliation{Theoretische Physik III, Ruhr-Universit\"at Bochum,
  D-44780 Bochum, Germany}
\author{Ilya M. Eremin}
\affiliation{Theoretische Physik III, Ruhr-Universit\"at Bochum,
  D-44780 Bochum, Germany}

\maketitle
\section{Correlated electronic structure approach}
The charge self-consistent~\cite{grieger12} combination of DFT, self-interaction correction
(SIC) and dynamical mean-field theory (DMFT), i.e. the so-called DFT+sicDMFT scheme~\cite{lechermann19}, is
put into practice. The Ni sites act as quantum impurities and Coulomb interactions on oxygen enter by SIC on
the pseudopotential level~\cite{korner10}. The DFT part consists of a mixed-basis
pseudopotential code~\cite{elsaesser90,lechermann02,mbpp_code} and SIC is applied to the O$(2s,2p)$ orbitals
via weight factors $w_p$. While the $2s$ orbital is fully corrected with $w_p=1.0$, the
choice~\cite{korner10,lechermann19,lechermann20-1} $w_p=0.8$ is used for $2p$ orbitals. Continuous-time
quantum Monte Carlo in hybridization-expansion scheme~\cite{werner06} as implemented in the
TRIQS code~\cite{parcollet15,seth16} solves the DMFT problem. A five-orbital Slater-Hamiltonian, parameterized
by Hubbard $U=10$\,eV and Hund exchange $J_{\rm H}=1$\,eV~\cite{lechermann20-1}, governs the correlated subspace
defined by Ni$(3d)$ projected-local orbitals~\cite{amadon08}, and the fully-localized-limit double-counting
scheme~\cite{anisimov93} is applied. In view of the wide Kohn-Sham projection space, encompassing Ni$(3d)$,
O$(2p)$ and partly La$(5d)$, the choice for the Hubbard $U$ and Hund $J_{\rm H}$ builds up on our
previous work on nickelates~\cite{lechermann20-1,lechermann22-2}, as well as on related GW+EDMFT
studies~\cite{choikutepov16,kanggwedmft25} for prototypical NiO. Crystallographic data are taken from
experiment~\cite{liu23}.

\section{Superconducting instability at strong coupling}
In this work, we implement the linearized gap equation as previously employed in the context of strong coupling with DFT+sicDMFT Green's functions \cite{nourafkan2016correlation,ryee24} 
\begin{align}
    \lambda(T) \Delta_{k, \eta_1 \eta_2} = -\left( \frac{k_B T}{2N_\mathbf{k}} \right) 
    \sum_{k', \eta_3 \ldots \eta_6} 
    \Gamma^{s/t}_{k, \eta_1 \eta_2; k', \eta_3 \eta_4}
     G_{\eta_3\eta_5}(k')G_{\eta_4\eta_6}(-k')
    \Delta_{k', \eta_5 \eta_6},
\end{align}
where $\eta = (l,\mu)$ is a combined index for sublattice (layer) and orbital degrees of freedom and $k = (\mathbf{k},i\omega_n)$ is the fermionic momentum-frequency four vector with Matsubara frequencies $\omega_n = (2n+1)\pi/T$. The above equation constitutes an eigenvalue problem with an eigenvalue $\lambda$ and the eigenvector $\Delta_{\eta_1\eta_2}(k)$ being the frequency dependent superconducting gap function. The square matrix is of size $N_{\mathbf{k}}\times2N_n\times N_{\mu}^2\times N_l^2$ with $N_\mathbf{k}$, $N_n$, $N_{\mu}$ and  $N_l$ denoting the number of momentum points, the number of positive Matsubara frequencies, the number of orbitals and the number of sublattices, respectively. The DFT+sicDMFT Green's functions are calculated on a $N_{\mathbf{k}} = 11^3$ momentum mesh ($9^3$ for pressurized La$_3$Ni$_2$O$_7$). Taking into account both Ni-3d$e_g$ orbitals for a bilayer structure implies $N_{\mu}=2$ and $N_l=2$. Owing to the substantial size of the matrix, the analysis incorporates only up to $N_n = 8$ Matsubara frequencies when both $e_g$ orbitals are considered. The calculation was also repeated neglecting the $d_{x^2-y^2}$ orbital and using $N_n = 16$ Matsubara frequencies. The matrix is based on the spin singlet/triplet pairing vertex $\Gamma^{s/t}$, which is computed herein employing the DFT+sicDMFT Green's functions $G_{\eta_1\eta_2}(k)$ with static screened irreducible vertices $U^{sp/ch}$ to the spin and charge susceptibilities. The bare susceptibility bubble is given by 
\begin{align}
    \chi^0_{\eta_1\eta_4,\eta_2\eta_3}(q) = -\frac{T}{N} \sum_{k}  G_{\eta_1\eta_2}(k+q) G_{\eta_3\eta_4}(k),
\label{Eq:nonintSus}
\end{align}
where $q = (\mathbf{q},i\nu_n)$ is a bosonic momentum-frequency four-vector with $\nu_n = 2n\pi/T$ denoting bosonic Matsubara frequencies. For the summation in Eq.~\ref{Eq:nonintSus}, a number of $300$ positive Matsubara frequency components are included. The RPA-like spin and charge susceptibilities are
\begin{equation}
	\label{Eq:RPAMatrixEqEvenOdd}
	\chi^{\rm sp/ch}_{\eta_1\eta_4,\eta_2\eta_3}
	=
	\left[
	\mathbb{1} \mp
	\chi^{0}
	\bar{U}^{\rm sp/ch}
	\right]^{-1}_{\eta_1\eta_4,\eta_5\eta_6}
    \chi^{0}_{\eta_5\eta_6,\eta_2\eta_3},
\end{equation}
where we assume local and frequency-independent irreducible spin and charge vertices $\bar{U}^{\rm sp/ch}$ depending on the effective screened multiorbital on-site Coulomb interactions, namely intraorbital Coulomb interaction $\bar{U}$, interorbital Coulomb interaction $\bar{U}'$, Hund's coupling $\bar{J}_{\rm H}$ and pair-hopping interaction $\bar{J}'$. This strategy was already employed in Ref.~\cite{nourafkan2016correlation}. The effective interactions are further assumed to be linked by spin rotational invariance $\bar{U}' = \bar{U} - 2\bar{J}_{\rm H}$ and $\bar{J}_{\rm H}=\bar{J}'$. In the case of hole-doped La$_3$Ni$_2$O$_6$, we employ effective interactions corresponding to a Stoner enhancement factor of $\alpha_{\rm sp} = 0.9$ for which the superconducting eigenvalue is already larger than unity. The Stoner enhancement factor describes the leading eigenvalue of $\chi^{0}\bar{U}^{\rm sp/ch}$ with unity corresponding to a magnetic transition according to the Stoner criterion. In the case of pressurized La$_3$Ni$_2$O$_7$, the used effective interactions correspond to a larger Stoner enhancement factor of $\alpha_{\rm sp} = 0.985$ for which the superconducting eigenvalue is almost unity. The exact eigenvalues depend on the number of Matsubara frequencies. For both systems, we fixed $\bar{J}_{\rm H}/\bar{U} = 0.1$. The spin(charge) irreducible vertex has an orbital dependence
\begin{align}
\bar{U}^{sp(ch)}_{\mu_2\mu_3;\mu_1\mu_4}  = 
	\begin{cases}
		\bar{U}  (\bar{U})  & \mu_1=\mu_2=\mu_3=\mu_4  \\
		\bar{U}' (2 \bar{J}_{\rm H} - \bar{U}') &  \mu_1=\mu_2\neq \mu_3=\mu_4  \\ 
		\bar{J}_{\rm H} (2 \bar{U}' - \bar{J}_{\rm H})  &\mu_1=\mu_4\neq \mu_2=\mu_3  \\
		\bar{J}' (\bar{J}')  &\mu_1=\mu_3\neq \mu_2=\mu_4 \\
		0 (0)  & {\rm otherwise}.  
	\end{cases}
\end{align}
Since on-site interactions are considered, only $l_1 = l_2 = l_3 = l_4$ sublattice components are non-zero. In bilayer systems neglecting interbilayer interactions \cite{yamase2024theory,boetzel24}, the physical measurable susceptibility can be written in terms of even and odd susceptibilities $\chi^{\rm even/odd} = \chi_\parallel \pm \chi_\perp$. The spin susceptibility is given by
\begin{align}
    \chi_{\text{spin}}(q) = \sum_{\mu_1,\mu_2} \left[ {\chi}^{\rm even}_{\mu_1,\mu_2} \cos^2(\frac{q_z d}{2}) + {\chi}^{\rm odd}_{\mu_1,\mu_2} \sin^2(\frac{q_z d}{2}) \right],
   \label{eq:SpinSusceptibility}
\end{align}
where the bilayer structure factors depend on the thickness of the bilayer $d$. The pairing vertices for zero-momentum spin singlet Cooper pairs are connected to the RPA susceptibilities via
\begin{align}
    \Gamma^{\rm s}_{k, \eta_1 \eta_2; k', \eta_3 \eta_4} = \Lambda^{\rm s}_{\eta_1\eta_3;\eta_4\eta_2}  +\frac{3}{2} [ \Psi^{\rm sp}_{\eta_2\eta_3;\eta_4\eta_1}(k-k') + 
    \Psi^{\rm sp}_{\eta_1\eta_3;\eta_4\eta_2}(k+k')  ] 
    -\frac{1}{2} [ \Psi^{\rm ch}_{\eta_2\eta_3;\eta_4\eta_1}(k-k')  +
    \Psi^{\rm ch}_{\eta_1\eta_3;\eta_4\eta_2}(k+k') ],
\end{align}    
with $\Psi^{\rm sp/ch}(q) = \bar{U}^{\rm sp/ch} \chi^{\rm sp/ch}(q) \bar{U}^{\rm sp/ch}$ and $\Lambda^{\rm s}$ being the bare spin singlet interaction given by $ \Lambda^{s}= 3/2 U^{\rm sp} + 1/2 U^{\rm ch}$. For the triplet pairing, the corresponding vertex is 
\begin{align}
    \Gamma^{\rm t}_{k, \eta_1 \eta_2; k', \eta_3 \eta_4} = \Lambda^{\rm t}_{\eta_1\eta_3;\eta_4\eta_2} 
    -\frac{1}{2} [ \Psi^{\rm sp}_{\eta_2\eta_3;\eta_4\eta_1}(k-k') - 
    \Psi^{\rm sp}_{\eta_1\eta_3;\eta_4\eta_2}(k+k')  ] 
    -\frac{1}{2} [ \Psi^{\rm ch}_{\eta_2\eta_3;\eta_4\eta_1}(k-k') -
    \Psi^{\rm ch}_{\eta_1\eta_3;\eta_4\eta_2}(k+k') ],
\end{align}
with $ \Lambda^{t}= - 1/2 U^{\rm sp} + 1/2 U^{\rm ch}$. Using $2N_n$ fermionic Matsubara frequencies in the eigenvalue problem necessitates to calculate $4N_n - 1$ bosonic Matsubara frequency susceptibility components. 

\subsection{Frequency and orbital dependence in hole-doped La$_3$Ni$_2$O$_6$}
\begin{figure*}[t]
      \includegraphics[width=\linewidth]{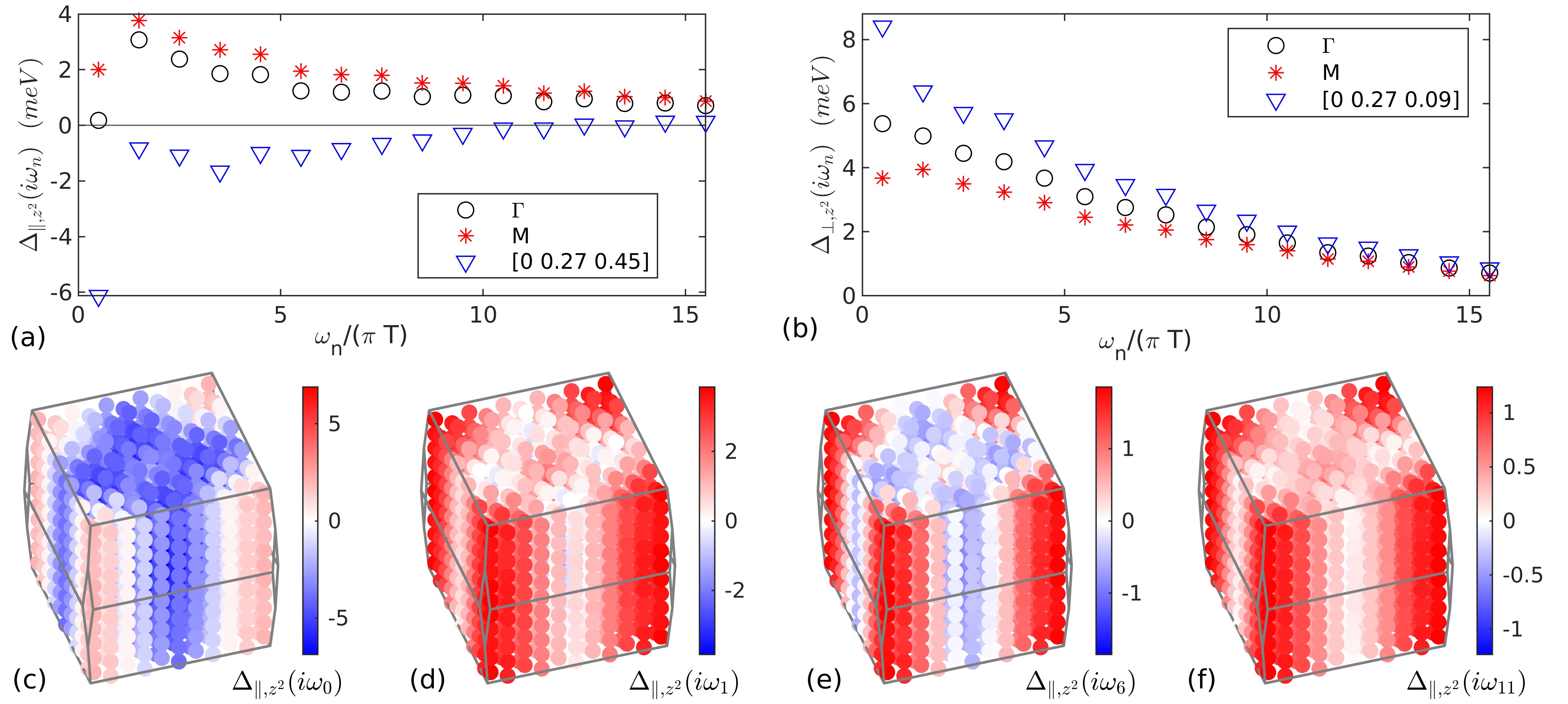}
      \caption{Matsubara frequency dependence of the calculated superconducting gap function (real part) for the Ni-$d_{z^2}$-orbital component in the first Brillouin Zone. In (a) and (b) the frequency dependence at representative momenta is shown for the intralayer and interlayer components, respectively. The full momentum dependence of the intralayer component is shown in the first Brillouin zone for selected Matsubara frequencies (c)-(f). 
}\label{figSM1}
\end{figure*}	
Within the weak coupling regime, the superconducting gap function exhibits a monotonic decrease as the amplitude of the Matsubara frequencies increase. For hole-doped La$_3$Ni$_2$O$_6$, we find a non-trivial dependence on Matsubara frequencies, which is shown in Fig.~\ref{figSM1} for the Ni-$3d_{z^2}$ orbital. The superconducting gap structure for the intralayer component in Fig.~\ref{figSM1}a significantly changes between the first and second Matsubara frequency for three selected momenta. Besides $\Gamma$ and $M$ point (pseudo-tetragonal notation), the momentum for which the highest gap amplitude is realized. Also, although not very obvious from the visualization, there is some not negligible variation in the $q_z$ direction, especially for smaller in-plane momenta. The frequency dependence is further visualized in the complete Brillouin Zone in Fig.~\ref{figSM1}c-f. The evolution from second to higher frequencies is still non-trivial but somewhat smoother than from first to second positive Matsubara frequency. In contrast, the interlayer component is almost monotonically decreasing. This behavior is consistent with naive expectations, since the interlayer component is not affected by Coulomb repulsion. The non-trivial behavior for the intralayer component reflects the sizable correlations of the system.

For the presented results with 16 positive Matsubara frequencies, the Ni-$3d_{x^2-y^2}$ orbital has been neglected. 
\begin{figure*}[t]
      \includegraphics[width=\linewidth]{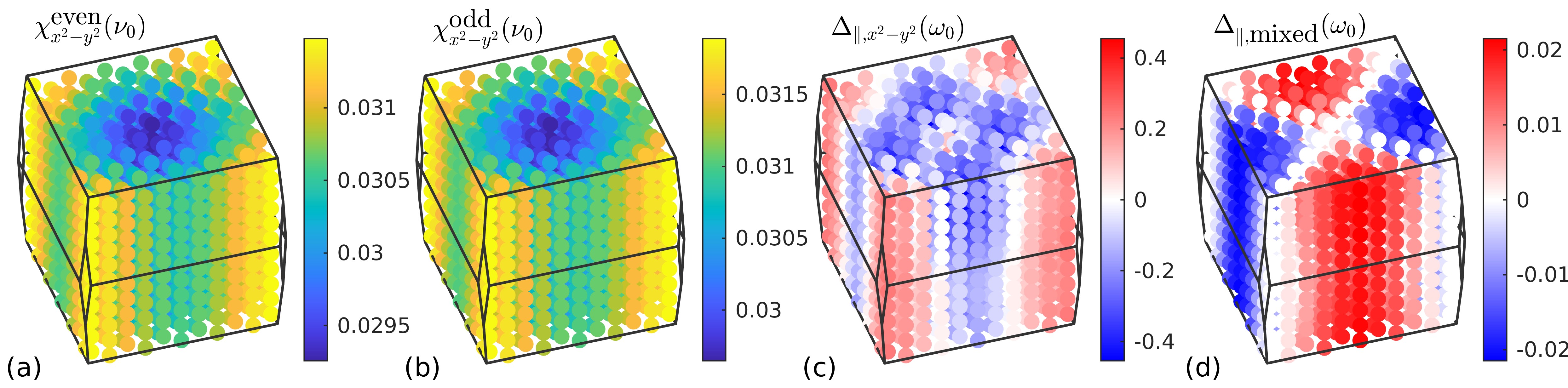}
      \caption{Spin susceptibility and superconducting gap function for the lowest positive Matsubara frequency for the Ni-$d_{x^2-y^2}$-orbital component in the first BZ. Note that the even susceptibility in (a) odd susceptibility in (b) are almost identical. The real part of the intralayer $d_{x^2-y^2}$ orbital and mixed orbital superconducting gap functions are shown in (c) and (d), respectively. Th susceptibility is given in units of states/$e$V, and the superconducting gaps are in units of meV.
}\label{figSM2}
\end{figure*}	
For justification, we show results for the Ni-$3d_{x^2-y^2}$ orbital in Fig.~\ref{figSM2}. Note that the even and odd spin susceptibility shown in Fig.~\ref{figSM2}a and Fig.~\ref{figSM2}b are more than one and a half order of magnitude smaller than the odd susceptibility shown in the main text. The same is true for all the mixed orbital components, which are not shown. For the Ni-$3d_{x^2-y^2}$ orbital, even and odd susceptibilities are almost identical, indicating $\chi_{\parallel,x^2-y^2} \gg \chi_{\perp,x^2-y^2}$. Further, the intralayer $d_{x^2-y^2}$ gap function shown in Fig.~\ref{figSM2}c precisely follows the gap structure of the $d_{z^2}$ component but is more than a magnitude smaller. This is also true for the higher frequency components. The interlayer $d_{x^2-y^2}$ gap function is smaller by several orders of magnitude. For completeness, the interorbital intralayer gap function is presented in Fig.~\ref{figSM2}d. Its $d_{x^2-y^2}$ symmetry is enforced by the symmetry of the interorbital Green's function due to the $d_{x^2-y^2}$ type hybridization of both $e_g$ orbitals combined with the $A_{1g}$ symmetry of the intraorbital components \cite{ryee24,botzel2024theory}. 

\subsection{Comparison with pressurized La$_3$Ni$_2$O$_7$}
To have a reference system, we also solved the linearized gap equation based on DFT+sicDMFT Green's function for pressurized La$_3$Ni$_2$O$_7$ at 80 K from Ref.~\cite{lechermann23}. There, both orbitals are strongly correlated without striking orbital selectivity. Due to the presence of the inner apical oxygen in La$_3$Ni$_2$O$_7$, strong superexchange involving the vertically connected Ni-$d_{z^2}$ orbitals is expected. Surprisingly, the leading eigenvalue approaches unity for a significant larger Stoner enhancement factor $\alpha_{\rm sp} \approx 0.985$. 
\begin{figure*}[t]
      \includegraphics[width=\linewidth]{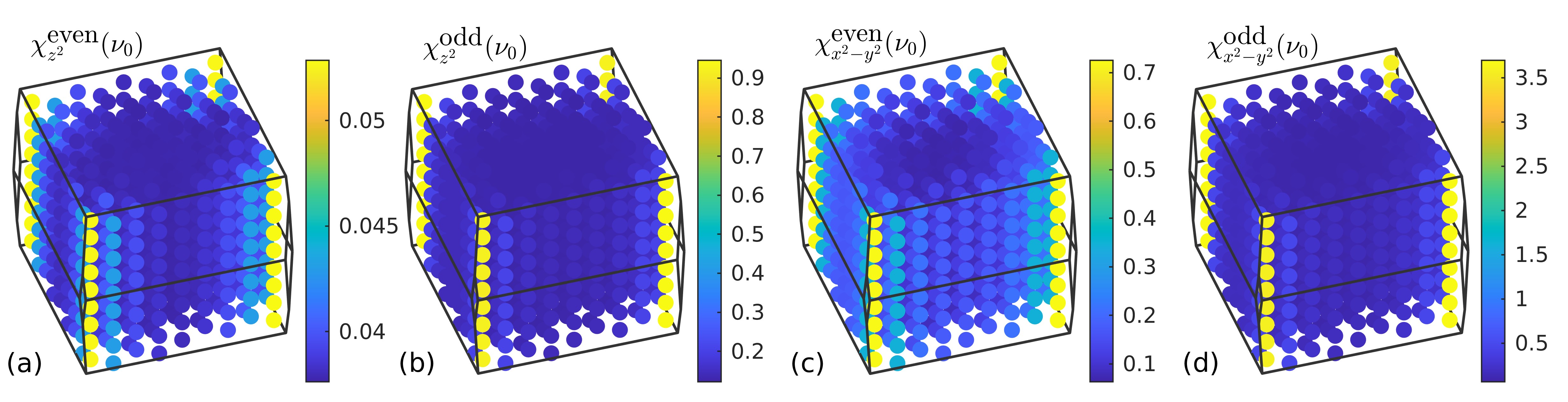}
      \caption{Spin susceptibility for the lowest positive Matsubara frequency in the first BZ for pressurized La$_3$Ni$_2$O$_7$. Even and odd Ni-$d_{z^2}$-orbital components are shown in (a) and (b), respectively. Even and odd Ni-$d_{x^2-y^2}$-orbital components are displayed in (c) and (d), respectively. Susceptibilities are in units of states/eV.
}\label{figSM3}
\end{figure*}	
The spin susceptibilities for the corresponding effective interactions for the Ni-$3d_{x^2-y^2}$ and Ni-$3d_{z^2}$ orbitals are shown in Fig.~\ref{figSM3}. The large differences of even and odd susceptibilities are reflecting strong interlayer effects. Opposite to hole-doped La$_3$Ni$_2$O$_6$, the Ni-$3d_{x^2-y^2}$ orbital contributions are larger than the Ni-$3d_{z^2}$ orbital contributions. Furthermore, note that the anisotropy of the susceptibility is substantially greater, with a very dominant peak at the $M$ point which is reminiscent of cuprates. However, the Ni-$3d_{z^2}$ orbital component still contributes significantly. Both, even and odd, susceptibilities exhibit negligible $q_z$ dependence. However, it should be noted that the physical measurable susceptibility is still dependent on $q_z$ due to the bilayer structure factors in Eq.~\ref{eq:SpinSusceptibility}. 

For the Stoner enhancement factor of $\alpha_{\rm sp} \approx 0.985$, there are several solutions of the eigenvalue problem with comparable eigenvalues. All these solutions have have non-trivial Matsubara frequency dependence and several important orbital and sublattice components. 
\begin{figure*}[t]
      \includegraphics[width=\linewidth]{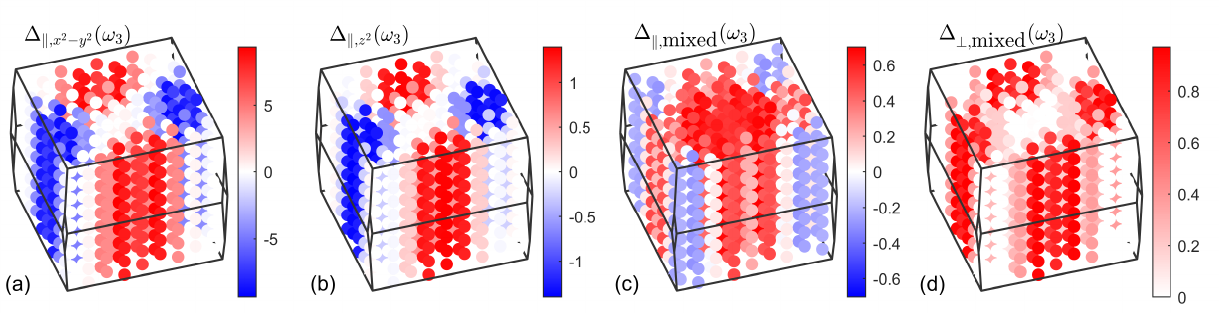}
      \caption{Gap structure for the $d_{x^2-y^2}$ intralayer solution in pressurized La$_3$Ni$_2$O$_7$ for the fourth positive Matsubara frequency component. The four largest orbital and sublattice components are shown, namely the intralayer $d_{x^2-y^2}$, $d_{z^2}$ and mixed orbital components in (a), (b) and (c), respectively, and the interlayer mixed orbital component shown in (d). The superconducting gap is in units of meV.
}\label{figSM4}
\end{figure*}	
The leading solution has an eigenvalue of $\lambda_1 \approx 0.84$ and a dominant $d_{x^2-y^2}$ intralayer gap structure, visualized in Fig.~\ref{figSM4}a. Nevertheless, there are other sizable components to be considered, including the $d_{z^2}$ and the mixed orbital intralayer components, shown in Fig.~\ref{figSM4}b and Fig.~\ref{figSM4}c, respectively. The latter has a $s_\pm$ structure, which can be as aforementioned linked to the extra $d_{x^2-y^2}$ form factor due to orbital hybridization. In fact, we find that a intraorbital $d_{x^2-y^2}$ gap structure always comes with an interorbital $s_\pm$ gap structure and vice versa in agreement with Refs.~\cite{ryee24,botzel2024theory}. This is also true for the interlayer mixed orbital component (Fig.~\ref{figSM4}d). The interlayer intraorbital components also have a $d_{x^2-y^2}$ gap structure and are of similar size as the shown interlayer component. Furthermore, there is a non-trivial Matsubara frequency dependence with the largest gap values being realized at the fourth positive Matsubara component $\omega_3$. However, the gap structures of the different orbital and sublattice components is similar for all Matsubara frequencies. 

\begin{figure*}[t]
      \includegraphics[width=\linewidth]{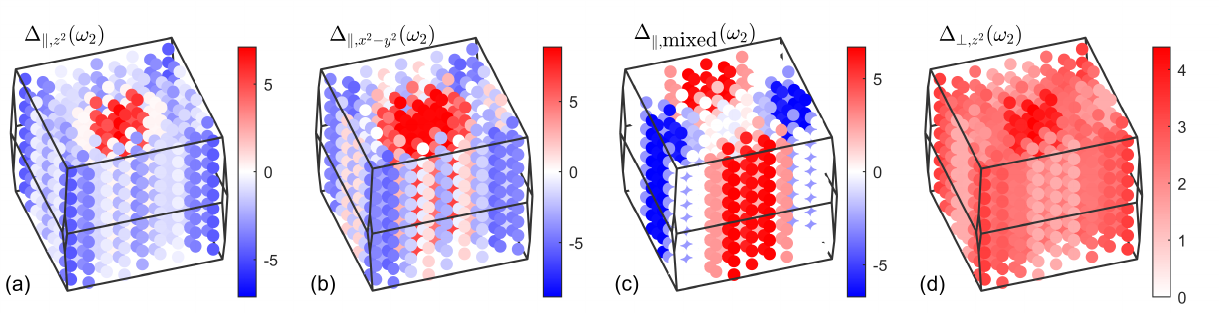}
      \caption{Gap structure for the extended $s_\pm$ solution in pressurized La$_3$Ni$_2$O$_7$ for the third positive Matsubara frequency component. The four largest orbital and sublattice components are shown, namely the intralayer $d_{z^2}$, $d_{x^2-y^2}$ and mixed orbital components in (a), (b) and (c), respectively, and the interlayer $d_{z^2}$ orbital component shown in (d). The superconducting gap is in units of meV.
}\label{figSM5}
\end{figure*}	

The subleading solution has an eigenvalue of $\lambda_2 \approx 0.62$ and also a dominant intralayer Ni-$3d_{x^2-y^2}$ component. Due to similarity to the leading solution, it is not shown. The main difference is that the intralayer Ni-$3d_{x^2-y^2}$ is more dominant and interlayer components are negligible, such that the subleading solution is more reminiscent of cuprates. The subsubleading solution has almost the same eigenvalue $\lambda_3 \approx 0.6$ and has a intraorbital $s_\pm$ gap structure with sizable interlayer component. For this solution, intralayer Ni-$3d_{z^2}$, Ni-$3d_{x^2-y^2}$ and mixed orbital and the interlayer Ni-$3d_{z^2}$ component shown in Fig.~\ref{figSM5}a-Fig.~\ref{figSM5}d are all on par. This solution also has a non-trivial Matsubara frequency dependence with the largest gap values being at the third positive Matsubara frequency $\omega_2$.

We also repeated the calculation for a lower Stoner enhancement factors. Moving away from the Stoner instability generally increases the Ni-$3d_{z^2}$ components compared to the Ni-$3d_{x^2-y^2}$ components. Below $\alpha_{\rm sp} \approx 0.9$ the intraorbital $s_\pm$ gap structure becomes leading. Therefore, in pressurized La$_3$Ni$_2$O$_7$ a close competition of several gap structures is found. Note that while the discussed gap symmetries are similar to the weak coupling regime \cite{lechermann23,zhang2024structural,yang2023possible}, the details of the gap structures in the strongly correlated regime are different. To be concrete, there are no signatures of the superconducting gaps following a well defined Fermi surface and there is a non-trivial Matsubara frequency dependence.

\bibliography{literatur}